\documentstyle[12pt]{article}

\def\be{\begin{eqnarray}}
\def\ee{\end{eqnarray}}

\def\partialleftright{\stackrel{\leftrightarrow}{\partial^\delta}}

\begin{document}
\begin{titlepage}

\begin{flushright}
CERN-TH/97-236\\
LPTHE-ORSAY 97/43  \\
hep-th/9709043 \\
\end{flushright}

\vskip.5cm
\begin{center}
{\huge{\bf Supersymmetry breaking in $M$-theory and quantization rules}}
\end{center}
\vskip1.5cm

\centerline{ Emilian Dudas $^{a,b}$ } 
\vskip 15pt
\centerline{$^{a}$ CERN-TH}
\centerline{CH-1211 Geneva 23, {\sc Switzerland}}
\vskip 3pt
\centerline{$^{b}$ Laboratoire de Physique Th\'eorique et Hautes Energies
\footnote{Laboratoire associ\'e au CNRS-URA-D0063.}}
\centerline{B\^at. 211, Univ. Paris-Sud, F-91405 Orsay Cedex, {\sc France}}
\vglue .5truecm

\begin{abstract}
\vskip 3pt
We analyze in detail supersymmetry breaking by compactification of the 
fifth dimension in M-theory in the compactification pattern 
$11d \rightarrow 5d \rightarrow 4d$ and find that a superpotential 
is generated for the 
complex fields coming from $5d$ hypermultiplets, namely the dilaton $S$ and
the complex structure moduli. Using general arguments it is shown that
these fields are always stabilized such that they don't contribute to
supersymmetry breaking, which is completely saturated by the K\"ahler 
moduli coming from vector multiplets.
  It is shown that this mechanism is the strong-coupling analog
of the Rohm-Witten quantization of the antisymmetric tensor field strength
of string theories. The effect of a gaugino condensate on one of the
boundaries is also considered.  	
  
\end{abstract}	

\vfill
\begin{flushleft}
CERN-TH/97-236\\
September 1997\\
\end{flushleft}

\end{titlepage} 
\section{Introduction}
It is largely believed nowdays that the strongly coupled regime of the 
heterotic string is  described within the
$M$-theory and, in particular, the strongly coupled 
$E_8 \times E_8$ heterotic string, traditionally considered as the most 
relevant one for 
phenomenology, can be
described, in the low energy limit, by the eleven-dimensional supergravity
with the two $E_8$ gauge factors living each on a 10-d boundary 
\cite{Horava&witten}.
The radius of the eleventh dimension is related to the string coupling by 
$R_{11} \sim 
\lambda_{st}^{2/3}$. 
So in the strongly coupled regime,
$R_{11}$ has to be large and possibly larger than the typical 
radius of the other six compact dimensions \cite{W}, \cite{BD}.
Describing four-dimensional physics from $E_8 \times E_8$ heterotic 
strongly coupled 
string should thus be equivalent to compactifing the eleven-dimensional 
supergravity on a Calabi-Yau manifold
and then compactifing the fifth dimension on $S^1/Z_2$.

Recently \cite{AQ1}, \cite{DG}, \cite{AQ2} (see also \cite{BRE} in the
type II context), attention was paid to the $4d$ supersymmetry 
breaking by the compactification from $5d$ to $4d$, by using the 
field-theoretical 
Scherk--Schwarz mechanism \cite{Scherk&Schwarz}. It was argued in \cite{DG} 
that the
results look like non-perturbative from the perturbative heterotic
string point of view. For the simplest truncation $11d \rightarrow 5d$ 
corresponding to no complex structure moduli,  a superpotential generation 
for $S$ was obtained.
The corresponding model has spontaneously broken supersymmetry with a zero
cosmological constant, the invariance $S \rightarrow 1/S$ and a minimum 
that is reached for $S=1$. We give in Section 2 the next non-trivial example
involving one complex structure modulus and then more general results are 
given by using generic arguments. Then the effect of a gaugino condensate
added on one of the boundaries is studied; the picture which emerges is
shown to be consistent.
  In Section 3 we show that the Scherk-Schwarz mechanism in this context 
describes in strong coupling regime the Rohm-Witten quantization 
\cite{RW} of the
antisymmetric tensor field strength \footnote{I would like to thank 
Jean-Pierre
Derendinger for discussions which lead me to study this analogy.} , 
used in the early days of string
phenomenology \cite{DIN} in the context of gaugino condensation for fixing 
the dilaton with a (tree-level) zero cosmological constant.
 

\section{The spontaneous breaking of $N=1$ supersymmetry in four dimensions 
by compactification}

To be specific, in the following we discuss the $5d \rightarrow 4d$
compactification.
The Scherk--Schwarz mechanism is a generalized dimensional reduction that 
allows for the 
fields a dependence in the compact coordinates. This 
dependence must satisfy some properties: 
it has to be in a factorizable form and has to correspond to an 
$R$-symmetry of the theory. If we denote generically by $\Phi$ all the
(boson and fermion) fields of the theory  , we have the following 
decomposition 
\be
\Phi (x,x_5) &=& U (x_5) \ \Phi (x) , \label{1}
\ee
where $x_5$ denotes the compact coordinate and $x$ non-compact ones.
This tensor decomposition is stable under product and exterior derivation.
For  $-\pi \le x_5 \le \pi$, defining
\be
\Phi (\pi)={\cal U} \Phi (-\pi) \ , \label{2}
\ee
consistency ask for ${\cal U}$ to be a symmetry of the $5d$ theory. 

This extended dimensional reduction 
generates a potential for the scalar fields corresponding to the kinetic
terms 
in the compact space. The requirement for this scalar potential to be 
positive imposes further restrictions on the form of $U$.
A solution was proposed by Scherk and Schwarz, by taking
\be
U &=& e^{M x_5}\ , \label{3}
\label{defUM}
\ee
where $M$ is an antihermitian matrix depending on the field representation.
When applied to the kinetic term for the $5d$ spin-${3\over 2}$ field, 
masses for the resulting $4d$ gravitinos are also generated.

The Horava-Witten projection $Z_2^{HW}$ acts as 
\be
Z_2^{HW} \phi (-x_5)=\eta \phi (x_5) \ , \ Z_2^{HW} \Psi (-x_5) = \eta 
\gamma_5 \Psi (x_5)
\ , \label{4}
\ee
where $\phi$ denote bosonic fields and $\Psi$ fermionic fields and $\eta=1$
for $g_{\mu \nu},g_{IJ},C_{5IJ},\\ C_{5 \mu \nu},\Psi_{\mu},\Psi_I,$ and 
$\eta=-1$ for $g_{5 \mu},C_{IJK},C_{\mu \nu \rho},\Psi_5$. 
The consistency of the Scherk-Schwarz mechanism with $Z_2^{HW}$ asks for
the condition \cite{kounnas}
\be
\{ Z_2^{HW} , M \} = 0 \ . \label{eq:s0}
\ee
It was proven in \cite{DG} that, at the field theory level, the only 
symmetry that is compatible
with the projection $Z_2^{HW}$  is a $U(1)$ 
subgroup of the $SU(2)_R$
symmetry appearing in the $N=2$ supersymmetry algebra. 
We quikly remind the results obtained within the
simplest truncation \cite{Witten85}, corresponding to an $SU(3)$ invariance
in the compactified space of volume $e^{3 \sigma}$ and $h_{(2,1)} =0$. 
In this case, in $5d$ the only matter
multiplet is the universal hypermultiplet $(e^{3 \sigma},C_{ijk}=\epsilon_{ijk}
a,C_{\mu \nu \rho})$, whose scalar fields parametrize
the coset ${{SU(2,1)} \over {SU(2) \times U(1)}}$ \cite{ferrara}. 
This structure can be simply viewed from $4d$ by a direct truncation 
({\it without} the $Z_2^{HW}$ projection). The lagrangian of the
universal hypermultiplet can be derived from the K\"ahler potential 
\cite{ferrara}
\be
{\cal K} =- \ln (S+S^{\dagger} - 2 a^{\dagger} a) \ , \label{eq:s1}
\ee
where the Hodge duality in $5d$ is 
\be
{\sqrt 2}e^{6 \sigma} G_{\mu\nu\rho\sigma} = \epsilon_{\mu\nu\rho\sigma\delta} 
( \partial^{\delta} a_1 + i a^\dagger \partialleftright a )
\label{s010}
\ee
and
$S=e^{3 \sigma}+a^{\dagger} a + i a_1$. The $SU(2)$
symmetry acts linearly on the redefined fields
\be
z_1 = {1-S \over 1+S} \ , \ z_2 = {2 a \over 1+S} \ , \label{eq:s2}
\ee
which form a doublet $(z_1, z_2)$. The $Z_2^{HW}$ projection acts as
$Z_2^{HW} S =S$, $Z_2^{HW} a = -a$, which translates on the $SU(2)$ doublet
in the obvious way:
\be 
Z_2^{HW}
\left( 
\begin{array}{c} 
z_1 \\ 
z_2 
\end{array} 
\right)  
&=& 
\left(
\begin{array}{cc} 
1 & 0 \\ 
0 & -1
\end{array}
\right)
\left(
\begin{array}{c}
z_1 \\
z_2
\end{array}
\right) \ . \label{eq:s3}
\ee
The Scherk--Schwarz decomposition in this case reads explicitly \cite{DG}
\be 
\left( 
\begin{array}{c} 
{\hat z}_1 \\ 
{\hat z}_2 
\end{array} 
\right)  
&=& 
\left(
\begin{array}{cc} 
\cos m x_5 & \sin m x_5 e^{i \theta} \\ 
-\sin m x_5 e^{-i \theta} & \cos m x_5
\end{array}
\right)
\left(
\begin{array}{c}
z_1 \\
z_2
\end{array}
\right) \ ,\label{eq:s4}
\ee
corresponding to the matrix defined in (\ref{defUM}) $M=im (cos \theta
\sigma_2 + sin \theta \sigma_1)$ and where $m$ is a real mass parameter
and $\theta$ a phase.\footnote{The possibility of adding this phase was 
noticed
in the second reference in \cite{AQ2}. However, this phase add no new 
physical
freedom and will be put to zero in the following.} 
Notice that, 
thanks to the anticommutation relation (\ref{eq:s0}), which is clearly 
verified, the fields ${\hat z}_i$ have
the same $Z_2^{HW}$ parities as the fields $z_i$.
The resulting scalar potential in $4d$ in Einstein metric is computed from 
the 
kinetic terms of the $({\hat z}_1, {\hat z}_2)$ fields derived from (\ref{eq:s1}). 
After making $z_2=0$, corresponding to the projection $Z_2^{HW}$, it is
easily worked out and  can be seen as a superpotential generation for $S$. 
The $4d$ theory is completely described by
\be
K &=&-\ln (S+S^{\dagger}) - 3 \ln (T+T^{\dagger}) \ , \ W= 2 m (1 + S) 
\ , \nonumber \\
V &=&{4 m^2 \over (S+S^{\dagger}) (T+T^{\dagger})^3} |1-S|^2 \ .  \label{eq:s5}
\ee
The resulting model is of a no-scale type \cite{CFKN}. This is a general
result for models obtained by the Scherk-Schwarz mechanism.
Notice that in the $z_1$ variable the superpotential is just a constant
$W=  2 {\sqrt 2} m$.
The minimum of the scalar potential is $S=1$ and gives a 
spontaneously broken supergravity model with a zero
cosmological constant. The order parameter for supersymmetry breaking
is the gravitino mass $m_{3/2}^{2} =e^{{\cal K}} |W|^2 = 
{8 m^2 / (T+T^{\dagger})^3}$.

The next non-trivial example corresponds to a compactification with $h_{(2,1)}
=1$, which can be obtained for example with  $Z_6$, $Z_{8}$ or $Z_{12}$ 
projections in the
compactified space. The $5d$ theory contains two hypermultiplets $(S,a_S)$,
with $C_{ijk}=\epsilon_{ijk}a_S$
and $(U,a_U)$, with  $C_{ij \bar k}=a_U$ and two vector multiplets.
The hypermultiplets scalars 
span the coset $U(2,2)/U(2) \times U(2)$. The corresponding K\"ahler 
potential is
\be
K=- \ln det ({\cal T} + {\cal T}^{\dagger} ) \ , where \
{\cal T} &=&
\left(
\begin{array}{cc}
S & 2 a_S \\
2 a_U & U
\end{array}
\right) \ . \label{s6}
\ee
The $SU(2)_R$ symmetry acts linearly on the matrix $z=z_0 (1-{\cal T}) 
(1+{\cal T})^{-1}$, in 
terms of which the K\"ahler potential is (up to a K\"ahler transformation)
$K=-Tr \ln (z_0 z_0^{\dagger} -z z^{\dagger})$. Here $z_0$ is an arbitrary, 
fixed matrix which, without loosing generality can be put to $z_0=1$ in the 
following. Imposing the Horava-Witten projection means
$a_S=a_U=0$ and the $4d$ theory contains the moduli $S,U$ coming from
$5d$ hypermultiplets and $T_1,T_2,T_3$ moduli, two coming from the two $5d$
vector multiplets and one (the overall volume) from the $5d$ gravitational
multiplet . The Scherk-Schwarz decomposition reads then
\be 
\hat z  
&=& 
\left(
\begin{array}{cc}
{1-S \over 1+S} & 0 \\
0 & {1-U \over 1+U}
\end{array}
\right)
\left(
\begin{array}{cc} 
\cos m x_5 & \sin m x_5 \\ 
-\sin m x_5 & \cos m x_5
\end{array}
\right) \ . \label{s8}
\ee
The $4d$ potential is computed as
\be
V=K_{ij, \bar k \bar l} \partial_5 {\hat z}^{ij} \partial^5 {\hat z}^{\dagger
\bar k \bar l} = m^2 e^{-3 \gamma} {|z_{11}|^2 + |z_{22}|^2 \over 
(1-|z_{11}|^2) (1- |z_{22}|^2) } \  \label{s80}
\ee
in $4d$ SUGRA units, where $g_{55}=e^{2 \gamma} \equiv t^2$ and 
$g_{\mu \nu}^{(5)}=e^{-\gamma} g_{\mu \nu}^{(4)}$. 
As in the previous example, the result coresponds to a superpotential 
generation $W=2 {\sqrt 2} m$. In the $S,U$ variables, the lagrangian is
\be
K &=& -\ln (S+S^{\dagger}) - \ln (U+U^{\dagger})- \sum_{i=1}^3 \ln 
(T_i+T_i^{\dagger}) \ , W= {\sqrt 2} m (1+S) (1+U) , \nonumber \\
V &=&{2 m^2 \over (S+S^{\dagger}) (U+U^{\dagger}) \prod_i (T_i+T_i^{\dagger})} 
\left[ |1-S|^2 |1+U|^2 + |1+S|^2 |1-U|^2 \right] \  . \label{s9}
\ee
The vacuum corresponds to $S=U=1$. 
It is easy to compute the physical masses and to check that 
$m_S^2=m_U^2=m_{3/2}^2$, in accordance with the Scherk-Schwarz mechanism.
Notice that the auxiliary fields $G_S=G_U=0$ don't contribute to supersymmetry 
breaking. If we integrate-in the field $U$ and put $T_i=T$ we recover the 
previous example (\ref{eq:s5}). This is valid for any generalization by
integrating-in the complex structure moduli $U_{\alpha}$. 

General results can be obtained by noticing that the K\"ahler moduli $T_i$ 
coming
from $5d$ vector multiplets are described by special K\"ahler geometry 
\cite{GST} and therefore their K\"ahler potential is
\be
K= -\ln {\cal F } \ \ , \ {\cal F}={1 \over 6} c^{ijk} (T_i+T_i^{\dagger}) 
(T_j+T_j^{\dagger}) (T_k+T_k^{\dagger}) \ , 
\label{s10}
\ee
where $c^{ijk}$ are the intersection numbers of the
Calabi-Yau manifold and ${\cal F}$ is related to the prepotential
$F={i \over 6} c^{ijk} T_i T_j T_k$ (for example, $F=iT^3$ for the first model
and $F=iT_1T_2T_3$ for the second model in this paragraph). On the other hand, 
as the $SU(2)_R$ symmetry
doesn't act on the vector moduli, there is no induced superpotential in $4d$
and so the whole vector moduli lagrangian is described by ${\cal F}$.
Then, by using $t_i {\cal F}^i = 3 {\cal F}$, where $2 t_i = T_i +
T_i^{\dagger}$, 
it is easy to show that \cite{FKZ} $G^i G_i = 3$ (where $G=K + \ln |W|^2$)
and the breaking of supersymmetry
is saturated by K\"ahler moduli. 
With these considerations, it is easy to determine the goldstino direction.
Indeed, by compactification, as the hypermultiplets don't contribute to
supersymmetry breaking, the goldstino should be a combination of $\Psi_5$
(fifth component of a Majorana gravitino)
and chiral fermions coming from $5d$ vector multiplets.
The $SU(2)_R$ acts explicitly on the last ones and they become massive.
On the other hand, due to the special structure of the $5d$ gravitino
kinetic term
\be
L_{kin}=- {1 \over 2} {\bar \Psi}_{\mu} {\Gamma}^{\mu \nu \rho} D_{\nu}
{\Psi}_{\rho} \ , \label{s100}
\ee  
the component $\Psi_5$ which survives after the Horava-Witten projection
cannot acquires a mass, therefore it must be identified with the goldstino.
\footnote{After our argument was completed, we learned that this was recently
noticed also in the second ref. in \cite{AQ2}.}

Using \cite{FKZ} $G^i = -2 t_i$ and denoting 
by $\chi_i$ the fermion associated to the moduli $T_i$, we get the expression 
for the goldstino
\be
g = e^{G \over 2} G_i \chi^i \sim \ -{m_{3/2} \over t^3} 
c^{ijk} t_i t_j \chi_k \sim \Psi_5 \ , \label{s11}
\ee
where $\Psi_5$ is, as we said, the fifth component of the gravitino, whose 
counterpart
$\Psi_{\mu}$ is projected by $Z_2^{HW}$. In order to compute (\ref{s11})
we used the fact that $G_{\alpha}=0$ for complex structure $U_{\alpha}$
fields. The last equality in (\ref{s11})
is just the supersymmetric partner of the relation 
$t^3 = {1/6}c^{ijk} t_i t_j t_k$,
where $t$ is the overall volume moduli coming from the $5d$ gravitational
multiplet $g_{55}=t^2$.
Therefore, for general models with complex 
structure moduli $U_{\alpha}$, zero cosmological constant (which is always
obtained at tree level in the Scherk-Schwarz mechanism) asks for $G_S =
G_{\alpha}=0$. This constraint, together with 
$m_S^2 = m_{\alpha}^2 = m_{3/2}^2$
can be used in order to construct effective lagrangians directly in $4d$.
One example of model obtained along these lines corresponds to
$h_{(1,1)}=h_{(2,1)}=3$. The lagrangian is a simple generalization of
(\ref{s9})
\be
K &=& -\ln (S+S^{\dagger}) - \sum_{\alpha=1}^3 \ln (U_{\alpha}+
U_{\alpha}^{\dagger})- \sum_{i=1}^3 \ln
(T_i+T_i^{\dagger}) \ , \\  
W &=& {m \over \sqrt 2} (1+S) \prod_{\alpha}(1+U_{\alpha}) \ . \label{s110}
\ee
In all the examples we worked out (including other $U$ moduli fields), {\it the
generated superpotential $W(S,U_{\alpha})$ is a constant in the variables 
where the $SU(2)_R$ symmetry acts linearly}, so we conjecture that this is a 
general result.

 An important question is of course the possible values the parameter $m$
can take. To answer this question, we use the result of Hull and Townsend
\cite{HT} which states that generically any global symmetry is broken to the
corresponding discrete subgroup by the Dirac charge quantization. In 
particular,
the $SU(2,R)_R$ symmetry should be broken to the $SL(2,Z)_R$ subgroup.
Consequently, we must impose the condition
\be {\cal U}
&=&
\left(
\begin{array}{cc}
cos 2 \pi m & sin 2 \pi m \\
-sin 2 \pi m & com 2 \pi m
\end{array}
\right) \ \in SL(2,Z) \ ,\label{s12}
\ee
which has as solutions $m= \pm 1/2 , \pm 1/4$ in units of $M_P$. 

It is useful to express the gravitino mass in a more usual way
by using relations obtained in \cite{DG}. We redefine $m =
n M_P$, $n$ being then a pure number. Then we get, for all the models
discussed above
\be
m_{3/2}^2 = {8 n^2 M_P^2 \over (T+T^{\dagger})^3} ={n^2 M_{11}^2 \over
t^2}= {n^2 \over \rho^2} \ , \label{s13}
\ee
where $\rho$ is the fifth radius and we used the relation \cite{DG}
$t M_{11}^2 = M_P^2$, which is the analog of the relation $s M_s^2 = 
M_P^2$ of the perturbative heterotic string\footnote{In \cite{DG} it
was considered the possibility that $m \sim M_{11}$, in which case the
gravitino mass was smaller compared to (\ref{s13}).}. The final expression 
(\ref{s13}) is of course the usual expression for Kaluza-Klein type 
masses. 
Notice that we can also rewrite the gravitino mass as
\be
m_{3/2} = n {M_{11}^3 \over M_P^2} \ , \label{131}
\ee
in a way that will become very transparent later on, when we will couple
a gaugino condensate on one boundary of the system.
As explained in \cite{AQ2}, the matter fields feel the supersymmetry
breaking only through radiative corrections and get soft masses which are
generically $m_{soft} \sim m_{3/2}^2/M_P$.

A next step in understanding the dynamics of the strongly coupled heterotic
string is the coupling of the five dimensional bulk to the
two $4d$ boundaries containing the matter and the gauge fields.\footnote{This
issue was recently considered from a complementary point of view in
\cite{NOY}, \cite{LT}.} 
In this paper, we neglect the problems (discussed in \cite{W}) due to the
modified Bianchi identity
\be
d G \sim \sum_i \delta (x_5-x_i) (tr F_i^2 - {1 \over 2} tr R^2 ) dx^{11}
\ , \label{s130}
\ee
where $x_i=0,\pi$ are the positions of the two boundaries,
because of our computational limitation in solving exactly the supersymmetry
conditions in the compactified space. While this is an important omission,
we believe including its effects will not qualitatively change our results.

We incorporate the effect of a gaugino condensate on one
(strongly coupled) boundary, let's say at $x_5=0$. It was noticed in
\cite{horava} that, in complete analogy with the weakly coupled case,
the $11d-10d$ lagrangian contains the perfect square
\be
{-1 \over 12 k_{11}^2} \int d^{11}x \sqrt{g} { \left[ 
G_{ABC11}-{\sqrt 2 \over 16 \pi} ({k_{11} \over 4 \pi})^{2/3} \delta (x^{11})
{\bar \lambda}^a \Gamma_{ABC} {\lambda}^a \right]}^2 \ , \label{s14}
\ee
where $k_{11}$ is the $11d$ gravitational constant and $\lambda^a$ are
gauginos living on the boundary. We consider only the simplest case $h_{(2,1)}
=0$. Then the compactification of the above term (\ref{s14}) to $5d$ 
changes the field strength $G_{5ijk}$. This corresponds to the shift in the
kinetic term for the universal complex field $a$ as
\be
\partial_5 a \rightarrow \partial_5 a - \beta \delta (x_5) {\bar \lambda 
\lambda}
\ , \label{s15}
\ee
where $\beta$ is related to the coefficient of the fermion bilinear in
(\ref{s14}) and ${\bar \lambda} \lambda$ is  the $4d$ gaugino
condensate. According to \cite{Horava&witten}, $\delta(x_5)$ is defined
here to transform as a scalar under diffeomorphisms. This means it
implicitly contain a vierbien $e_5^5=e^{\gamma}$ in its definition;
also, the $11$ index in $G_{ABC11}$ is a Lorentz one and for doing the
computation it must be changed into an Einstein index, a
fact which turns out to be crucial in the following.
The rest of the $5d$ lagrangian is unchanged and we can 
apply again the Scherk-Schwarz mechanism. 
The scalar potential compared to the case without gaugino condensate
is shifted according to
\be
V \rightarrow V-\beta e^{-\gamma} [e^{-3 \hat \sigma} \partial_5 ({\hat a}+
{\hat a}^{\dagger})]_{(x_5=0)}
{\bar \lambda} \lambda + \beta^2 e^{-3 \sigma} \delta (0) ({{\bar \lambda}
\lambda})^2  \ , \label{s160}
\ee
where $e^{-\gamma}$ factor originates from the inverse vierbein necessary
to convert a Lorentz index into an Einstein one, as discussed above.
Using (\ref{eq:s4}) for doing the computation and then
going into $4d$ SUGRA ($g_{\mu \nu}^{(5)}=e^{-\gamma} g_{\mu \nu}^{(4)}$ and
$\lambda \rightarrow e^{3 \gamma \over 4}\lambda $) units we get the result
\be
V= {1 \over (S+S^{\dagger})} \left[ 4m^2 {|1-S|^2 \over
(T+T^{\dagger})^3} -
{2 \sqrt 2 m \beta \over (T+T^{\dagger})^{3/2}} (2-S-S^{\dagger}) 
{\bar \lambda \lambda} + 
{2 \beta^2 } \delta (0) ({\bar \lambda \lambda})^2 \right] \ .
\label{s16}
\ee
The last term in (\ref{s16}) is of course ill-defined and can hardly help
us in finding the correct result. On the other hand, the first two terms have 
a clear tendency of forming a perfect square and therefore the correct result
should be
\be
V= {1 \over (S+S^{\dagger})} |{2m(1-S) \over (T+T^{\dagger})^{3/2}}-
{\sqrt 2 \beta}
{\bar \lambda \lambda}|^2 \ . \label{s17}
\ee
Puting back the mass units in our result, we see that we found (for
$m \sim M_P$, as we argued earlier)
\be
<{\bar \lambda \lambda}> \sim {M_P^3 \over t^{3/2}} = M_{11}^3 \ , 
\label{s170}
\ee
where in the last step we used again the formula $t M_{11}^2 = M_P^2$.
We therefore found explicitly the result conjectured in \cite{AQ2},
namely the gaugino condensation scale is the M-theory scale $M_{11}$.
Now the formulae (\ref{131}),(\ref{s170}) are of course simply explained 
in perfect
analogy with the gaugino condensation scenario in SUGRA \cite{Nilles}
\be
m_{3/2} \sim {<{\bar \lambda \lambda}> \over M_P^2} \ . \label{s171}
\ee 
In the presence of the condensate,
the vev of $S$ is shifted. Notice that, here $S$ is the volume
of compactified space  on the boundary 
containing the condensate and therefore is related to the gauge coupling of
the strongly coupled hidden sector. We now compute the volume of the 
compactified space on the other (observable) boundary. The result is
\be
{\cal V}(\pi) \equiv {\hat S}(\pi)= {{\cal V}(0) \over |cos^2 {m \pi \over 2} +
sin^2 {m \pi \over 2} S|^2} \ . \label{s18}
\ee
In the absence of the gaugino condensate, $S=1$ and therefore ${\cal V}(\pi)
={\cal V}(0)$, the two boundaries being perfectly symmetric. 
In the presence of the gaugino condensate, $S <1$ and
we get ${\cal V}(\pi) > {\cal V}(0)$ such that the observable world has a 
smaller gauge coupling.
The picture is therefore consistent and reminds us the situation described
in \cite{W}, where it was shown that, due to the modified Bianchi identity
(\ref{s130}) a similar phenomenon occurs. 

    Finally, we note that, if we don't impose the Horava-Witten
projection, the resulting N=2 model in $4d$ has interesting properties,
too. Namely, the Scherk-Schwarz masses, given by the
parameter $m$ which becomes now the N=2 central charge, are BPS saturated
and in consequence the tree level result is actually exact. Indeed, the $5d$ 
supersymmetry algebra becomes in $4d$, in a Weyl notation 
\be
\left(
\begin{array}{cc}
\{Q_2,Q_1\}  & \{Q_2,{\bar Q_2}\} \\
\{{\bar Q_1},Q_1\} & \{{\bar Q_1},{\bar Q_2}\}
\end{array}
\right) \
&=&
2 \left(
\begin{array}{cc}
P_5 & \sigma^{\mu} P_{\mu} \\
{\bar \sigma}^{\mu} P_{\mu} & -P_5
\end{array}
\right) \ ,\label{s28}
\ee
where $Q_1,Q_2$ are the two supersymmetry charges in $4d$ and $P_5$ is the
fifth momentum, which is therefore the central charge. The usual argument
gives here for the mass operator $M^2 \ge P_5^2$. It is now straightforward
to check the BPS relation $M^2=P_5^2$ by using Scherk-Schwarz decompositions
of type  (\ref{eq:s4}) and the spectrum of masses computed from the scalar
potential.

\section{Quantization of $S$ and of the complex structure moduli.}

It was shown \cite{RW} in the context of string theory that the field 
strength of the
antisymmetric tensor $B$, $H=d B - {\alpha' \over 2} \omega_{3Y}$ satisfies
the quantization rule
\be
\int_{C_3} H = {2 \pi \over T_2} p \ , \label{q1}
\ee
where $T_2$ is the string tension, $C_3$ is a closed three-manifold and $p$ 
is an integer. In particular, this
predicted in components $H_{ijk}=c \epsilon_{ijk}$, with $c$ a quantized
parameter which was used in \cite{DIN} in the gaugino condensation scenario
in order to break supersymmetry with a stabilized dilaton and zero
cosmological constant. 
Rohm and Witten argued that $c$ depends generically
on the complex structure moduli. 
 
 We claim now that similar quantization rules appear in our version of the
Scherk-Schwarz mechanism with an interpretation very similar in spirit to
that of Rohm and Witten.
More precisely, we show by an explicit computation that despite the fact
that the three form components $C_{ijk},C_{ij \bar k}, C_{\mu \nu \rho},\mu,
\nu,\\ \rho= 1,2,3,4$ are odd under $Z_2^{HW}$ and therefore have
no zero modes in $4d$, the corresponding field strengths have background
values leading to quantization rules.
The generalization of (\ref{q1}) in strong coupling regime is
\be
\int_{{S^1 / Z_2} \times C_3} G = {2 \pi \over T_3} p \ , \label{q3}
\ee
where $G=6 dC + a \delta (x^5) d x^5 \omega_{3Y}$ ($a$ is given in 
\cite{Horava&witten}) and $T_3$ is the membrane
tension. We neglect in this paper the Chern-Simmons possible contribution
to $G$. Their consequence can be discussed along the lines of ref. \cite{RW},
but this is beyond our goal here.
 In components, the field strength which interest us are
$G_{5ijk}, G_{5ij \bar k}$.
The components $C_{ijk}$ and $C_{ij \bar k}$
depend in a non-trivial way on $x^5$, so  we get the result
\be
\int dx^5  {\hat G}_{5ijk} = \epsilon_{ijk} [ {\hat a}(\pi)-
{\hat a}(-\pi) ] = 2 \epsilon_{ijk} {\hat a} (\pi) \ , \label{q4}
\ee
exactly because of the twist in the boundary conditions asked by the 
Scherk-Schwarz mechanism. This is equivalent with the presence of a five-brane
as a magnetic source for $G$.
Similar quantization rules appear from $G_{5ij \bar k}$.
The right-hand side of (\ref{q4}) can be easily calculated by using the results
of the preceding paragraph. The result is
\be  
{\hat G}_{5ijk}=3 m \epsilon_{ijk} {(S-1) (cos^2 {mx_5 \over 2} -sin^2 
{mx_5 \over 2} S) \over 
 (cos^2 {mx_5 \over 2} + sin^2 {mx_5 \over 2} S)^2} \ , \ 
{\hat a}(\pi) = {S-1 \over ctg {m \pi \over 2} +tg {m \pi \over 2} S} \ 
\  \label{q5}
\ee
for $h_{(2,1)}=0$ and
\be
({\hat G_S})_{5ijk}=3 m \epsilon_{ijk} {(S-1)(1+U) (cos^2 {mx_5 \over 2} 
-sin^2 {mx_5 \over 2} SU) \over
 (cos^2 {mx_5 \over 2} + sin^2 {mx_5 \over 2} SU)^2} \ , \nonumber \\ 
({\hat G_U})_{5ijk}=3 m \epsilon_{ijk} {(1+S)(U-1) (cos^2 {mx_5 \over 2}
-sin^2 {mx_5 \over 2} SU) \over
 (cos^2 {mx_5 \over 2} + sin^2 {mx_5 \over 2} SU)^2} \ , \nonumber \\
2 {\hat a_S}(\pi) = {(S-1)(1+U) \over ctg{m \pi \over 2} +tg {m \pi
\over 2}SU}  \ , 
2 {\hat a_U}(\pi) = {(1+S)(U-1) \over ctg{m \pi \over 2} +tg {m \pi
\over 2}SU}  \  \label{q6}
\ee
for $h_{(2,1)}=1$. For $U=1$ the second set of quantization rules (\ref{q6})
coincide with (\ref{q5}), as it should. As promised, the quantization 
conditions
involve explicitly the complex structure moduli and the dilaton $S$. We 
stress that the parameter $m$ was already quantized before, so 
the quantized quantities are really the vev's of the moduli fields.
Notice that at the global minimum values, the quanta of charge are zero. 
The quantization conditions tell us that the
classical minimum is valid at a non-perturbative level. Another, more
intriguing possibility, would to get a quantum minimum which satisfy the
quantization rules with a large $S$, in order to accomodate the low energy
observed phenomenology. 

  Another quantity of interest which appears here is a background value of
$G_{\mu \nu \rho \sigma}$\footnote{I would like to thank Pierre Bin\'etruy
for collaboration in this part of the paper.}. For simplicity reasons, we are 
computing it in the most simple case, $h_{(2,1)}=0$. 
We use (\ref{eq:s4}) in order to find  
\be
{\hat S} = {1- cos mx_5 z_1 \over 1 + cos mx_5 z_1} \ ,
{\hat a} = {- sin mx_5 z_1 \over 1 + cos mx_5 z_1} \ ,
{\hat a_1} = i { (z_1 - z_1^{\dagger}) cos mx_5  \over 
|1 + cos mx_5 z_1|^2 } \ . \label{q7}
\ee
Then, the Hodge duality (\ref{s010}) gives us the remarkably simple
result
\be
{\hat G}_{\mu \nu \rho \sigma}= -{m \over \sqrt 2} e^{-3 \sigma} a_1 sin mx_5 
\epsilon_{\mu \nu \rho \sigma} \ . \label{q8}
\ee
 The $4d$ lagrangian, neglecting for the moment the Scherk-Schwarz potential,
is invariant under discrete shifts of the axion field $a_1$. This means, by
using (\ref{q8}) that, on the $x_5=\pi$ boundary, $G_{\mu \nu \rho \sigma}$
is quantized too. It is already known that, in any case, $G$ is quantized
\cite{WQ} due to the modified Bianchi identity (\ref{s130}), so we showed
that this holds true in the context discussed here. In complete analogy 
with the discussion concerning $G_{5ijk}, G_{5ij \bar k}$, the scalar
potential (\ref{eq:s5}) reach its minimum for $a_1=0$,  corresponding
to a zero quantum of charge.

\section{Conclusions}
This paper studies a class of $4d$ models obtained by compactifying from
$11d \rightarrow 5d \rightarrow 4d$ the M-theory of Horava and Witten and
twisting the boundary conditions in the fifth dimension \`a la
Scherk-Schwarz. General features of such models are given, based on
geometrical properties of the $5d$ theory.  The presence of a gaugino
condensate on one of the boundaries is also included, showing that indeed
this corresponds to a larger gauge coupling on this boundary, still keeping
the cosmological constant to zero. We showed explicitly that $M_{11}$ is
the scale of the gaugino condensation and that $m_{3/2} \sim {< \bar \lambda
\lambda> / M_P^2}$, as conjectured in \cite{AQ2}. 

It is shown that, due to the twisted boundary conditions, magnetic charges
appear in the system giving quantization rules similar to that discussed
by Rohm and Witten in the case of perturbative string theories. We claim
that the Scherk-Schwarz mechanism here is a manifestation in strong coupling
regime of the Rohm-Witten mechanism. The addition of the gaugino condensate
leads to a physical picture which is very close to that discussed in
\cite{DIN}. More precisely, our picture is not that the Scherk-Schwarz
mechanism is the gaugino condensation in the M-theory context, but that
the gaugino condensation combine with the Scherk-Schwarz mechanism in the
same way the gaugino condensation combines with the quantization of the
antisymmetric field strength in the perturbative strings. 
On the other hand, in the M-theory case, the phenomenological perspectives
are certainly better, since we naturally get a condensation scale of order
$M_{11}$, compared to a scale of order $M_P$ in the perturbative case,
asked by the Rohm-Witten quantization rules. 

\section*{Acknowledgements}
It is a pleasure to thank I. Antoniadis, P. Bin\'etruy, J.P. Derendinger, 
C. Grojean, C. Kounnas, J. Mourad and M. Quiros for helpful discussions 
and comments.
  



\newpage
\begin{thebibliography}{99}

\bibitem{Horava&witten} P. Ho\v{r}ava and E. Witten,  {\it Nucl. Phys.}
{\bf B460} (1996) 506 and  {\bf B475} (1996) 94.

\bibitem{W} E. Witten, {\it Nucl. Phys.} {\bf B471} (1996) 135.

\bibitem{BD} T. Banks and M. Dine, {\it Nucl. Phys. }{\bf B479} (1996) 173,
hep-th/9609046; \\
E. Caceres, V.S. Kaplunovsky and I.M. Mandelberg, {\it Nucl. Phys.}
{\bf B493} (1997) 73; \\
E. Dudas and J. Mourad, {\it Phys. Lett.} {\bf B400} (1997) 71;\\
T. Li, J.L. Lopez and D.V. Nanopoulos,  hep-ph/9702237, hep-ph/9704247.

\bibitem{AQ1} I. Antoniadis and M. Quiros, {\it Phys. Lett. }{\bf B392}
(1997) 61.

\bibitem{DG} E. Dudas and C. Grojean, hep-th/9704177, {\it Nucl. Phys.} 
{\bf B}, to appear. 

\bibitem{AQ2} I. Antoniadis and M. Quiros, hep-th/9705037 and hep-th/9707208.

\bibitem{BRE} E. Bergshoeff, M. de Roo and E. Eyras, hep-th/9707130.

\bibitem{Scherk&Schwarz} J. Scherk and J.H. Schwarz,  {\it Nucl. Phys.}
{\bf B153} (1979) 61, and  {\it Phys. Lett.} {\bf B82}  (1979) 60;\\
E. Cremmer, J. Scherk and J.H. Schwarz,  {\it Phys. Lett.}  {\bf B84} (1979)
83;\\
P. Fayet, {\it Phys. Lett.} {\bf B159} (1985) 121, {\it Nucl. Phys.}
{\bf B263} (1986) 649.

\bibitem{RW} R. Rohm and E. Witten, {\it Ann. Phys.} {\bf 170} (1986) 454.

\bibitem{DIN} J.P. Derendinger, L.E. Ib\'a\~nez and H.P. Nilles, 
{\it Phys. Lett.} {\bf B155} (1985) 467;\\
M. Dine, R. Rohm, N. Seiberg and E. Witten, {\it Phys. Lett.} {\bf B156}
(1985) 55.

\bibitem{kounnas} C. Kounnas and M. Porrati, {\it Nucl. Phys. }{\bf B310}
(1988) 355; \\
S. Ferrara, C. Kounnas, M. Porrati and F. Zwirner, {\it Nucl. Phys.}
{\bf B318} (1989) 75.

\bibitem{Witten85} E. Witten, {\it Phys. Lett. } {\bf B155 }(1985) 151.

\bibitem{ferrara} S. Ferrara and S. Sabharwal, {\it Class. Quantum Grav.}
{\bf 6} (1989) L77, {\it Nucl. Phys.} {\bf B332} (1990) 317.

\bibitem{CFKN} E. Cremmer, S. Ferrara, C. Kounnas and D.V. Nanopoulos,
{\it Phys. Lett.} {\bf B133} (1983) 61;\\
J. Ellis, C. Kounnas and D.V. Nanopoulos, {\it Nucl. Phys.} {\bf B241}
(1984) 406 and {\bf B247} (1984) 373.

\bibitem{GST} M. G\"unaydin, G. Sierra and P.K. Townsend, {\it Nucl.
Phys.} {\bf B242} (1984) 244 , {\it Nucl. Phys. } {\bf B253} (1985) 573;\\
E. Cremmer, C. Kounnas, A. van Proeyen, J.P. Derendinger, S. Ferrara,
B. de Wit and L. Girardello, {\it Nucl. Phys. }{\bf B250} (1985) 385.

\bibitem{FKZ} S. Ferrara, C. Kounnas and F. Zwirner, {\it Nucl. Phys.}
{\bf B429} (1994) 589. .

\bibitem{HT} C.M. Hull and P.K. Townsend, {\it Nucl. Phys.} {\bf B438}
(1995) 409.

\bibitem{horava} P. Ho\v{r}ava, {\it Phys. Rev. }{\bf D54} (1996) 7561.

\bibitem{Nilles} S. Ferrara, L. Girardello and H.P. Nilles,
{\it Phys. Lett.} {\bf B125} (1983) 457.

\bibitem{NOY} H.P. Nilles, M. Olechowski and M. Yamaguchi, hep-th/9707143.

\bibitem{LT} Z. Lalak and S. Thomas, hep-th/9707223.

\bibitem{WQ} E. Witten, hep-th/9609122.

\end{thebibliography}
\end{document}